\documentclass{article}
\usepackage{latexsym, amsfonts, amssymb, graphics}
\usepackage{author}

\begin{document}

\title{Macroscopic determinism in noninteracting systems using large deviation theory}
\author{Brian R. La Cour \thanks{Corresponding author. E-mail address:
        blacour@physics.utexas.edu} \\
        William C. Schieve \\
        Ilya Prigogine Center for Studies in \\
        Statistical Mechanics and Complex Systems \\
        The University of Texas at Austin \\
        Austin, Texas 78712}
\maketitle

\begin{abstract}
We consider a general system of $n$ noninteracting identical particles which evolve under a given dynamical law and whose initial microstates are \textit{a priori\/} independent. The time evolution of the $n$-particle average of a bounded function on the particle microstates is then examined in the large $n$ limit. Using the theory of large deviations, we show that if the initial macroscopic average is constrained to be near a given value, $y$, then the macroscopic average at time $t$ converges in probability as $n \rightarrow \infty$ to a value, $\psi_t(y)$, given explicitly in terms of a canonical expectation.  Some general features of the graph of $\psi_t(y)$ versus $t$ are examined, particularly in regard to continuity, symmetry, and convergence.
\end{abstract}


%%%%%%%%%%%%%%%%%%%%%%%%%%%%%%%%%%%%%%%%%%%%%%%%%%%%%%%%%%%%%%%%%%%%%%%%%%%%

\section{Introduction}
\label{sect:Introduction}

The emergence of determinism in the macroscopic variables of a system from its underlying microscopic dynamics has been a subject of great importance in the field of statistical mechanics.  If one supposes the microscopic variables are rigidly deterministic, then the time evolution of the macroscopic variables as a function of the initial \emph{microstate} is of course rigidly deterministic as well.  Considered as a function of the initial \emph{macrostate}, however, its time evolution fails to be deterministic due to the multiplicity of microstates consistent with the given macrostate. This leads to the familiar ensemble description, as described by a conditional \textit{a priori\/} measure.  Given the highly irregular behavior of many macroscopic systems on the microscopic level, however, it may appear that all deterministic behavior is utterly lost on the macroscopic level.

By contrast, many macroscopic systems are described quite well by deterministic models, even if they are not deterministic in the strict sense.  Well known examples include thermal conduction, diffusion, hydrodynamics, and chemical reactions. Characteristic of many such systems is the presence of extensive macroscopic variables which are sums or averages over a great many microscopic quantities.  The task of deriving differential equations for such variables has been taken up by several researchers \cite{vanKampen1961,Kurtz1972,Kubo1973}, and here Markov processes have played an important role. In this approach, one derives a coupled set of differential equations for the moments of the macroscopic variable, where the deterministic behavior is given by the first moment when the dispersion goes to zero.  Since few physical observables are truly Markovian, delicate scaling between vanishing interactions and dilating time scales must be used to obtain an asymptotically Markovian process \cite{Spohn,Davies}.  However, the consistency of such assumptions with the underlying microscopic dynamics and the relevance of strong interactions have been repeatedly called into question \cite{Sklar,Mehra1972}. In particular, the relaxation of macroscopic systems to a state of equilibrium may seem inconsistent with the underlying reversible microscopic dynamics.

We take a somewhat different and more general view upon this problem. Instead of attempting to derive differential equations of motion for a class of macroscopic variables, we consider instead the existence and character of an emergent form of determinism for large systems, which we call macroscopic determinism. By this we mean that if the macrostate is initially constrained to be near a given value, $y$, then there exists a map $\psi_t$ such that the probability that the macrostate is near $\psi_t(y)$ at time $t$ approaches one as the number of particles approaches infinity.  For simplicity, we restrict attention to systems of dynamically noninteracting particles and macroscopic variables which take the form of  averages over these microscopic quantities. The mathematical theory of large deviations, an extension of the law of large numbers, provides an useful tool for addressing such questions and is used in \Sec \ref{sec:LDTA} to obtain the main result.  Our primary contribution has been to apply well-known equilibrium results to systems initially out of equilibrium.

The collection of macrostates $\set{\psi_t(y) : t \in T}$ for a given $y$ and set of times $T$ constitutes a collection of highly probable states, which we call the deterministic curve, akin to the concentration curve of P. and T. Ehrenfest in their discussion of Boltzmann's H-theorem \cite{Ehrenfest}. In \Sec \ref{sec:DC} we show that the macroscopic average converges to this curve in probability for any given finite, and in some cases infinite, set of times, though large deviations from this curve will persist whenever the number of particles is finite. For certain reversible microscopic dynamics which preserve the \textit{a priori\/} measure, the deterministic curve of a restricted class of macroscopioc variables is symmetric in time about the initial specification of the macroscopic variable. Furthermore, if the microscopic dynamical law is mixing, then the deterministic curve converges as $t \rightarrow \pm\infty$ to a value independent of $y$, a situation corresponding to equilibration of the macrostate.

%%%%%%%%%%%%%%%%%%%%%%%%%%%%%%%%%%%%%%%%%%%%%%%%%%%%%%%%%%%%%%%%%%%%%%%%%%%%

\section{Problem Description}
\label{sec:PD}

We begin with a mathematical description of the relevant physical quantities. The microstate space of a single particle is denoted by $X$, while the microstate space of the total $n$-particle system is given by the Cartesian product $X^n = X \times \stackrel{n}{\cdots} \times X$. The projection map $\pi_i$ is defined such that, if $(x_1, \ldots, x_n) \in X^n$ is the microstate of the system, then $\pi_i(x_1, \ldots, x_n) = x_i$ is the microstate of the $i^{\mathrm{th}}$ particle.  To apply the techniques of large deviation theory we shall need to make the mild assumption that $(X, d)$ is a Polish space, i.e.\ a complete separable metric space, with respect to some given metric $d$, and denote by $\mathcal{A}$ the set of Borel subsets of $X$ generated by the metric topology.

Let $\mathcal{P}(X)$ denote the set of Borel probability measures on $X$ and note that $\mathcal{P}(X)$ is itself a Polish space under the Prohorov metric $\rho$ induced by the metric $d$ on $X$ \cite[p.\ 317]{Dudley}. Denote by $\mu \in \mathcal{P}(X)$ the \textit{a priori\/} probability measure for the initial microstate of a given particle. In other words, for $C \in \mathcal{A}$, $\mu[C]$ is the probability that a given particle's initial microstate is in $C \subseteq X$, \emph{before} any conditioning on the macrostate has taken place. All particle microstates therefore have equal \textit{a priori\/} probability in the sense that they are uniformly distributed with respect to $\mu$.  Usually, $\mu$ is taken to be an invariant measure. The system microstates are assumed to be \textit{a priori} independent and identically distributed (i.i.d.\/) with marginal $\mu$. Thus, the \textit{a priori} distribution on $X^n$ is given by the product measure $\mu^n = \mu \times \stackrel{n}{\cdots} \times \mu$.

Let $g: X \rightarrow Y$ be a measurable function which is bounded and continuous $\mu$-almost everywhere (a.e.\/).  For a given particle microstate $x_i \in X$, $g(x_i)$ gives the corresponding single particle macrostate. For a collection of $n$ particles, the macroscopic average $G: X^n \rightarrow Y$ of $g$ is given by
\begin{equation}
G := \frac{1}{n} \sum_{i=1}^{n} g \circ \pi_i.
\end{equation}
It is to this macroscopic variable that we will focus our attention.  We shall take $Y$ to be a set of real numbers equipped with the Euclidean norm. To accommodate all possible values of $G$ as well as any accumulation points, we shall assume $Y = [\ymin,\; \ymax]$, where $\ymin = \inf g(X)$, $\ymax = \sup g(X)$, and $g(X)$ is the image of $X$ under $g$. Note that $G$ is indeed a measurable function since it is a finite sum of measurable functions.

The dynamics of the system are described by a family of measurable transformations $\Phi_{t}$ on $X^n$ indexed by the time parameter $t \in T \subseteq \mathbb{R}$, where $\Phi_{0}$ is the identity map. The macroscopic average at time $t$ is thus given by $G_{t} := G \circ \Phi_{t}$. We shall suppose that the particles are dynamically independent and identical in the sense that
\begin{equation}
\Phi_{t} = (\varphi_t \circ \pi_1, \ldots, \varphi_t \circ \pi_n),
\end{equation}
where $\varphi_t$ is a measurable transformation on $X$.  We shall further suppose that $\varphi_t$ is continuous $\mu$-a.e.\ on $X$ for each $t \in T$, though not necessarily continuous on $T$ for $\mu$-a.e.\ $x \in X$.

The macrostate $G_{t}$ may be considered a sample mean over the \textit{a priori\/} i.i.d.\ random variables $\set{\varphi_t \circ \pi_i}_{i=1}^{n}$ with common marginal $\mu \circ (g \circ \varphi_t)^{-1}$. Since $g$ is bounded, the weak law of large numbers implies $G_{t}$ converges in probability to the expectation value $\int_{X} g \circ \varphi_t \;\d \mu$ as $n \rightarrow \infty$; in other words,
\begin{equation}
\lim_{n \rightarrow \infty} \mu^{n}[G_{t} \in B] =
  \left\{
  \begin{array}{ll}
    0, & \mbox{if $\int_{X} g \circ \varphi_t \;\d \mu \not\in \closure{B}$,} \\
    1, & \mbox{if $\int_{X} g \circ \varphi_t \;\d \mu \in \interior{B}$} \\
  \end{array}
  \right.
\end{equation}
for any Borel set $B \subseteq Y$, where $\interior{B}$ is the interior of $B$ and $\closure{B}$ is its closure.  No limit is specified for points on the boundary, $\boundary{B}$, of $B$.

If the initial microstates are restricted so as to satisfy some initial macroscopic constraint, then the variables $\set{\varphi_t \circ \pi_i}_{i=1}^{n}$ may no longer be independent due to the correlations imposed by this conditioning.  A direct application of the law of large numbers is therefore no longer valid.  Suppose, in particular, that the initial macrostate $G_{0} = G$ is constrained to be in a small region $B_{\delta}$ containing a given macrostate $y$. We will show that $G_{t}$ converges in conditional probability to some $\psi_t(y)$; in other words,
\begin{equation}
\lim_{n \rightarrow \infty} \mu^{n}[G_{t} \in B \:| G \in B_{\delta}] =
  \left\{
  \begin{array}{ll}
    0, & \mbox{if $\psi_t(y) \not\in \closure{B}$,} \\
    1, & \mbox{if $\psi_t(y) \in \interior{B}$}. \\
  \end{array}
  \right.
\end{equation}
where $\psi_t(y)$ is the expectation value of $g \circ \varphi_t$ with respect to a new probability measure determined by $y$.

In \Sec \ref{sec:LLNA} we shall consider three values of $y$, namely $\ymin, \ymax$, and $y_* = \int_{X} g \;\d \mu$, for which the law of large numbers alone may be used to determine $\psi_t(y)$ and prove convergence in conditional probability. Later, in \Sec \ref{sec:LDTA}, convergence is proven for $y \in (\ymin, \ymax) = \interior{Y}$ using the theory of large deviations, from which a general expression for $\psi_t(y)$ is obtained in terms of the expectation of $g \circ \varphi_t$ with respect to a suitable canonical measure.

%%%%%%%%%%%%%%%%%%%%%%%%%%%%%%%%%%%%%%%%%%%%%%%%%%%%%%%%%%%%%%%%%%%%%%%%%%%%

\section{Law of Large Numbers Approach}
\label{sec:LLNA}

We consider first an approach using only the law of large numbers, which will be valid for $y = y_* := \int_{X} g \;\d \mu$ and, in certain cases, $y = \ymin$ and $y=\ymax$.  We begin with the latter two cases.

% Extremal Initial Macrostates

Suppose $\mu[\set{g = \ymin}] > 0$ and note that, since $\set{ G = \ymin } = \set{ g = \ymin } \times \stackrel{n}{\cdots} \times \set{ g = \ymin }$,
\begin{equation}
\mu^{n}\left[\left.A_1 \times \cdots \times A_n \;\right| \set{G = \ymin}\right] = \prod_{i=1}^{n} \mu[A_i \:| \set{g = \ymin}],
\end{equation}
for any $A_1, \ldots, A_n$ in $\mathcal{A}$.  Thus, conditioned on $\set{G = \ymin}$, $G_{t}$ is a sample mean of i.i.d.\ random variables with common marginal $\mu[(g \circ \varphi_{t})^{-1}(\;\cdot\;) \:| \set{g = \ymin}]$.  The weak law of large numbers then implies
\begin{equation}
\lim_{n \rightarrow \infty} \mu^{n}[G_{t} \in B \:|
\set{G = \ymin}] =
  \left\{
  \begin{array}{ll}
    0, & \mbox{if $\psi_{t}(\ymin) \not\in \closure{B}$,} \\
    1, & \mbox{if $\psi_{t}(\ymin) \in \interior{B}$} \\
  \end{array}
  \right.
\end{equation}
for any Borel set $B \subseteq Y$, where
\begin{equation}
\label{eqn:ECM} % Deterministic Curve for Minimum
\psi_{t}(\ymin) := \int_{X} g \circ \varphi_{t} \;\d \mu[\;\cdot\; \:| \set{g = \ymin}].
\end{equation}
A similar result holds for conditioning on $\set{G = \ymax}$, provided $\mu[\set{g = \ymax}] > 0$.

% Equilibrium Initial Macrostates

Let us turn now to the case $y = y_*$, where we suppose $y_* \in \interior{Y}$. Given a set $B_{\delta} \subseteq Y$ whose interior contains $y_*$, we wish to examine the limit
\begin{equation}
\lim_{n \rightarrow \infty} \mu^n[\set{G_{t} \in B} \:| \set{G \in B_{\delta}}],
\end{equation}
for an arbitrary Borel subset $B$ of $Y$.

To parallel the large deviation approach of the following section, we consider the convergence of the empirical measure $L_n$, defined by
\begin{equation}
L_n(x_1, \ldots, x_n) := \frac{1}{n} \sum_{i=1}^{n} \delta_{x_i},
\end{equation}
where $\delta_{x_i}$ is the unit point measure on $x_i$. Given $(x_1, \ldots, x_n) \in X^n$ and $C \subseteq X$ measurable, $L_n(x_1, \ldots, x_n) \in \mathcal{P}(X)$, and $L_n(x_1, \ldots, x_n)[C]$ is the fraction of particles whose microstate is in $C$.  A useful result is that $G_{t}$, and hence $G$, may be written in terms of $L_n$, since
\begin{equation}
G_{t}(x_1, \ldots, x_n) = \frac{1}{n} \sum_{i=1}^{n} g(\varphi_t(x_i)) = \int_X g \circ \varphi_t \;\d L_n(x_1, \ldots, x_n).
\end{equation}
Using the above relation, convergence properties of $G_{t}$ may then be deduced from those of $L_n$.

Now, since $(X,d)$ is a separable space, $L_n$ converges almost surely to $\mu$ \cite[p.\ 313]{Dudley} and hence in probability as well. Thus, for any Borel set $A \subseteq \mathcal{P}(X)$,
\begin{equation}
\label{eqn:EMCP} % Empirical Measure Convergence in Probability
  \lim_{n \rightarrow \infty} \mu^n[\set{L_n \in A}] =
  \left\{
  \begin{array}{ll}
    0, & \mbox{if $\mu \not\in \closure{A}$,} \\
    1, & \mbox{if $\mu \in \interior{A}$.} \\
  \end{array}
  \right.
\end{equation}
Now suppose $A_{\delta}$ is a Borel subset of $\mathcal{P}(X)$ such that $\mu \in \interior{A_{\delta}}$.  Using \Eqn (\ref{eqn:EMCP}), we have that
\begin{equation}
\label{eqn:EMCPCE} % Empirical Measure Convergence in Probability, Conditioned on Equilibrium
  \lim_{n \rightarrow \infty} \mu^n[\set{L_n \in A} \:| \set{L_n \in A_{\delta}}] =
  \left\{
  \begin{array}{ll}
    0, & \mbox{if $\mu \not\in \closure{A}$,} \\
    1,& \mbox{if $\mu \in \interior{A}$,} \\
  \end{array}
  \right.
\end{equation}
since $\mu^n[\set{L_n \in A_{\delta}}] \rightarrow 1$ as $n \rightarrow \infty$.

To apply this result to the convergence of $G_{t}$, we define the expectation function $E_{g}(P) := \int_X g \;\d P$ for $P \in \mathcal{P}(X)$ and note that
\begin{equation}
E_{g} \circ L_n = \frac{1}{n} \sum_{i=1}^{n} g \circ \pi_i = G.
\end{equation}
Since $g$ is bounded and continuous $\mu$-a.e.\, $E_{g}$ is continuous at $\mu$ in the Prohorov metric topology \cite{Dudley,Ash}. (Of course, if $P \prec \mu$, then $E_{g}$ is continuous at $P$ as well.) Thus, $E_{g}(\mu) \in \interior{B}$ implies $\mu \in \interior{E_{g}^{-1}(B)}$, and similarly $E_{g}(\mu) \not\in \closure{B}$ implies $\mu \not\in \closure{E_{g}^{-1}(B)}$.

Now, since $g \circ \varphi_t$ is also bounded and continuous $\mu$-a.e.\/, $\psi_t(y_*) = E_{g \circ \varphi_t}(\mu) \in \interior{B}$ implies $\mu \in \interior{E_{g \circ \varphi_t}^{-1}(B)}$, and similarly $\psi_t(y_*) \not\in \closure{B}$ implies $\mu \not\in \closure{E_{g \circ \varphi_t}^{-1}(B)}$. Setting $A = E_{g \circ \varphi_t}^{-1}(B)$ and $A_{\delta} = E_{g}^{-1}(B_{\delta})$ in \Eqn (\ref{eqn:EMCPCE}), we conclude that
\begin{equation}
\label{eqn:MACPCE} % Macroscopic Average Convergence in Probability, Conditioned on Equilibrium
  \lim_{n \rightarrow \infty} \mu^n[\set{G_{t} \in B}
  \:| \set{G \in B_{\delta}}] =
  \left\{
  \begin{array}{ll}
    0, & \mbox{if $\psi_t(y_*) \not\in \closure{B}$,} \\
    1, & \mbox{if $\psi_t(y_*) \in \interior{B}$,} \\
  \end{array}
  \right.
\end{equation}
where
\begin{equation}
\label{eqn:ECE} % Deterministic Curve for Equilibrium
\psi_t(y_*) := E_{g \circ \varphi_t}(\mu) = \int g \circ \varphi_t \;\d \mu.
\end{equation}
If $\mu$ is an invariant measure, i.e.\ $\mu \circ \varphi_t^{-1} = \mu$, we obtain the rather trivial result that $G_{t}$ converges in probability to its equilibrium value, $y_*$, as $n \rightarrow \infty$.

The general problem, wherein $y \in \interior{Y} \setminus \set{y_*}$, cannot be addressed with the law of large numbers alone.  To see this, consider a set $B_{\delta} \subset \interior{Y}$ such that $y_* \not\in \closure{B_{\delta}}$ and let $B \subseteq Y$ be some Borel set. We wish to evaluate the following conditional probability as $n \rightarrow \infty$:
\[
\mu^n[\set{G_{t} \in B} \:| \set{G \in
B_{\delta}}] = \mu^n[\set{L_n \in E_{g \circ \varphi_t}^{-1}(B)} \:| \set{L_n \in E_{g}^{-1}(B_{\delta})}]
\]
\[
= \mu^n[\set{L_n \in E_{g \circ \varphi_t}^{-1}(B) \cap
E_{g}^{-1}(B_{\delta})}] \;/\; \mu^n[\set{L_n \in E_{g}^{-1}(B_{\delta})}].
\]
Since $y_* = E_{g}(\mu) \not\in \closure{B_{\delta}}$, $\mu \not\in \closure{E_{g}^{-1}(B_{\delta})}$ and the denominator goes to zero. Since $\mu \not\in \closure{E_{g \circ \varphi_t}^{-1}(B)} \cap \closure{E_{g}^{-1}(B_{\delta})} \supseteq \closure{E_{g \circ \varphi_t}^{-1}(B) \cap E_{g}^{-1}(B_{\delta})}$, we see that the numerator goes to zero as well, leaving the limit indeterminant. To proceed further requires more detailed information about the rate of convergence of each limit, which the law of large numbers alone cannot provide.  For this we turn to the theory of large deviations.

%%%%%%%%%%%%%%%%%%%%%%%%%%%%%%%%%%%%%%%%%%%%%%%%%%%%%%%%%%%%%%%%%%%%%%%%%%%%

\section{Large Deviation Theory Approach}
\label{sec:LDTA}

We have seen in the previous section that the weak law of large numbers is insufficient for evaluating limiting conditional probabilities when the initial macrostate is not $y_*$. (More precisely, this is true when we condition on a \emph{set} of macrostates whose closure does not contain $y_*$.) Large deviation theory provides a tool for evaluating such limits and, through its application, provides an explicit expression for $\psi_t(y)$ even when $y \neq y_*$.  This approach is a refinement of the weak law of large numbers for cases in which the probabilities converge exponentially fast at a rate given in terms of a function $I$, the so-called ``rate function.''  The ground work for this theory was established by Boltzmann \cite{Boltzmann1877} in his study of the asymptotic properties of multinomials and later applied by Einstein \cite{Einstein1907} in his analysis of fluctuations. Recent years have seen great development of this relatively new field of mathematical probability, including applications in equilibrium statistical mechanics, stochastic processes, and mathematical statistics \cite{Deuschel_and_Stroock,Ellis,Dembo_and_Zeitouni,Dupuis_and_Ellis}.  Ruelle \cite{Ruelle} and Lanford \cite{Lanford1973} have developed similar techniques for studying the equilibrium distributions of dynamical maps, where the (negative) Kolmogorov-Sinai entropy serves as a rate function \cite{Kifer1990}.

In \Sec \ref{ssec:LDT} we define rate functions and the large deviation principle.  The main result is \Thm \ref{thm:LDCC} regarding convergence in probability for conditional probabilities.  In \Sec \ref{ssec:GC} we take up the notion of Gibbs Conditioning, which will allow us to apply \Thm \ref{thm:LDCC} to cases in which we condition on an arbitrary initial macrostate $y$.  Our results are similar to those of \cite{Dembo_and_Zeitouni}, who use the stronger $\tau$-topology on $\mathcal{P}(X)$, and follow from an approach adapted from \cite{Ellis1999}, who considers discrete macrostates.

%---------------------------------------------------------------------------

\subsection{Large Deviation Theory}
\label{ssec:LDT}

For a topological space $(X, \mathcal{T})$, a function $I$ mapping $X$ into $[0, \infty]$ is called a \emph{rate function\/} iff $\inf I(X) = 0$ and $I$ is lower semicontinuous, i.e.\ the preimage $I^{-1}([0, \alpha])$ is closed for all $\alpha \in [0, \infty)$. A \emph{good\/} rate function is one in which these preimages are compact.  The property of being lower semicontinous guarantees that $I$ attains its infimum on any compact set, from which it follows that a good rate function attains its infimum over any closed set \cite[pp.\ 4, 308]{Dembo_and_Zeitouni}.  Any point, $x_*$, at which $I(x_*) = 0$ is called an \emph{equilibrium point\/}, since it corresponds to a state of maximum probability.

A sequence $\seq{P_n}_{n \in \mathbb{N}}$ of probability measures on the Borel subsets of $X$ is said to satisfy a \emph{large deviation principle} with rate function $I$ iff there exists a sequence $\seq{a_{n}}_{n \in \mathbb{N}}$ of positive numbers tending to infinity such that, for any Borel set $A \subseteq X$,
\begin{equation}
\label{eqn:LDP} % Large Deviation Principle
-\inf I(\interior{A}) \le \liminf_{n \rightarrow \infty} \frac{1}{a_{n}} \log P_n[A]
\le \limsup_{n \rightarrow \infty} \frac{1}{a_{n}} \log P_n[A] \le - \inf I(\closure{A}).
\end{equation}
A set $A$ for which $\inf I(\interior{A}) = \inf I(\closure{A})$ is called an \emph{$I$-continuity set\/}.  Clearly for such sets we have
\begin{equation}
\label{eqn:LDPCS} % Large Deviation Principle for Continuity Sets
\lim_{n \rightarrow \infty} \frac{1}{a_{n}} \log P_{n}[A] = -\inf I(A).
\end{equation}
This result should be compared with the Boltzmann relation, $S = k_{\mathrm{B}} \log W$, relating the thermodynamic entropy, $S$, to a volume, $W$, in phase space, where $k_{\mathrm{B}}$ is Boltzmann's constant. The quantity $-a_n I$ serves as a negative entropy, while $P_{n}$ may be viewed as a normalized volume measure.

If $0 < \inf I(A) < \infty$, \Eqn(\ref{eqn:LDPCS}) implies that $P_{n}[A]$ converges to zero at least exponentially fast.  Specifically, from \Eqn(\ref{eqn:LDP}) we may deduce the following:  Given any Borel set $A \subseteq X$ and $\varepsilon > 0$, then for all $n$ sufficiently large,
\begin{equation}
\label{eqn:LDULB}  % Large Deviation Upper and Lower Bounds
\exp[-a_{n} (1+\varepsilon) \inf I(\interior{A})] \le P_{n}[A] \le
\exp[-a_{n} (1-\varepsilon) \inf I(\closure{A})],
\end{equation}
provided\/ $\inf I(\interior{A}) > 0$ and\/ $\inf I(\closure{A}) < \infty$. Equality for the lower bound may hold only if\/ $\inf I(\interior{A}) = \infty$, where\/ $\e^{-\infty} := 0$, while equality for the upper bound may hold only if\/ $\inf I(\closure{A}) = 0$.

If, for example, there is a unique equilibrium point $x_{*}$, then $x_{*} \not\in \closure{A}$ implies $\inf I(\closure{A}) > 0$. Provided $\inf I(\closure{A}) < \infty$ and taking $\varepsilon = 1/2$, say, this implies $P_{n}[A] \le \exp[-n \inf I(\closure{A})/2]$ for all $n$ sufficiently large. If $\inf I(\closure{A}) = \infty$, then $A$ is an $I$-continuity set and \Eqn (\ref{eqn:LDPCS}) implies $a_{n}^{-1} \log P_{n}[A] \rightarrow -\infty $. Thus, if $x_* \not\in \closure{A}$, then $P_{n}[A] \rightarrow 0$ as $n \rightarrow \infty$, and we recover the weak law of large numbers, i.e.\
\begin{equation}
\label{eqn:LDZOC}  % Large Deviation Zero-One Convergence
\lim_{n \rightarrow \infty} P_{n}[A] = \left\{
\begin{array}{ll}
  0, & \mbox{if\/ $x_* \not\in \closure{A}$}, \\
  1, & \mbox{if\/ $x_* \in \interior{A}$},
\end{array} \right.
\end{equation}
for any Borel set $A \subseteq X$.

For our purposes, we would like to show a result similar to \Eqn~(\ref{eqn:LDZOC}) for the case in which we condition on a suitable initial condition.  In particular, we would like to obtain a result analogous to \Eqn (\ref{eqn:EMCPCE}). The following theorem, stated in its general form, may be used for this purpose. The proof is deferred to \App \ref{sec:LDCC}.

\begin{thm}
\label{thm:LDCC}  % Large Deviation Conditional Convergence
Suppose\/ $\seq{P_{n}}_{n \in \mathbb{N}}$ satisfies a large deviation principle with a good rate function\/ $I$ and a unique equilibrium point\/ $x_*$.  Let\/ $B \subseteq X$ be an $I$-continuity set such that\/ $x_* \not\in \boundary{B}$ and there exists a unique\/ $x_{B} \in \closure{B}$ such that\/ $I(x_{B}) = \inf I(\closure{B}) < \infty$. Given any Borel set\/ $A \subseteq X$,
\begin{equation}
\label{eqn:LDCC} % Large Deviation Conditional Convergence
  \lim_{n \rightarrow \infty} P_{n}[A \:| B] =
  \left\{
  \begin{array}{ll}
    0, & \mbox{if\/ $x_{B} \not\in \closure{A}$,} \\
    1, & \mbox{if\/ $x_{B} \in \interior{A}$.}
  \end{array}
  \right.
\end{equation}
\end{thm}

In the next section, we shall consider a large deviation principle for the sequence $\seq{\mu^n \circ L_n^{-1}}_{n \in \mathbb{N}}$ of distributions of empirical measures.  Conditioning on $A_{\delta} = E_{g}^{-1}(B_{\delta})$ will then give rise to a new equilibrium probability measure, $P_{\lambda}$, in general different from $\mu$.  We shall show that under these conditions \Thm \ref{thm:LDCC} is satisfied, thus establishing convergence of $L_{n}$ in conditional probability to $P_{\lambda}$.

%---------------------------------------------------------------------------

\subsection{Gibbs Conditioning}
\label{ssec:GC}

In his 1877 paper, Boltzmann proved that the asymptotically most probable configuration for a gas of $n$ particles with a finite number of macrostates is given by a multinomial distribution. Sanov's theorem \cite[p.\ 70]{Deuschel_and_Stroock}, a modern refinement of this classic result, states that the sequence $\seq{\mu^n \circ L_n^{-1}}_{n \in \mathbb{N}}$ of distributions of empirical measures satisfies a large deviation principle with rate function $I_{\mu}: \mathcal{P}(X) \rightarrow [0, \infty]$. Here $I_{\mu}(P)$ is the (negative) Gibbs entropy of $P$ with respect to $\mu$, defined by
\begin{equation}
I_{\mu}(P) := \left\{
\begin{array}{ll}
 \int_{X}   \frac{\d P}{\d \mu} \log \frac{\d P}{\d \mu} \;\d \mu, & \mbox{if $P \prec \mu$,} \\
 \infty, & \mbox{otherwise}, \\
\end{array}
\right.
\end{equation}
where $0 \log 0 := 0$. It can be shown that $I_{\mu}$ is a good, strictly convex, rate function \cite[p.\ 240]{Dembo_and_Zeitouni} which attains its infimum uniquely at $\mu$ \cite[pp.\ 32--34]{Dupuis_and_Ellis}.

Given $y \in \interior{Y}$ and $\delta > 0$, let $A_{\delta} = E_{g}^{-1}(B_{\delta})$, where, $B_{\delta} = (y - \delta, y]$ when $y < y_*$, $B_{\delta} = [y, y + \delta)$ when $y > y_*$, and $B_{\delta} = (y_* - \delta, y_* + \delta)$ when $y = y_*$. (Note that $\mu \not\in \boundary{A_{\delta}}$, since $y_* \not\in \boundary{B_{\delta}}$ and $E_{g}$ is continuous at $\mu$.) We wish to consider the asymptotic behavior of the conditional probability $(\mu^{n} \circ L_{n}^{-1})[A \:| A_{\delta}] = \mu^{n}[\set{L_{n} \in A} \:| \set{L_{n} \in A_{\delta}}]$, as $n \rightarrow \infty$, for any Borel set $A \subseteq \mathcal{P}(X)$. For example, if $G$ is the macroscopic average energy of the system, then $\mu^n [\;\cdot\; \:| \set{G \in B_{\delta}}]$ is the microcanonical distribution on the ``thickened'' energy shell with energy $y$.

We will show that, conditioned on $\set{G \in B_{\delta}}$, the empirical measures $\seq{L_n}_{n \in \mathbb{N}}$ converge in probability to the canonical Gibbs measure $P_{\lambda}$, where $\lambda$ satisfies the constraint $y = \int_{X} g \;\d P_{\lambda}$ and
\begin{equation}
\d P_{\lambda}(x) := \frac{\e^{\lambda g(x)}}{Z(\lambda)} \;\d \mu(x).
\end{equation}
The normalization factor $Z(\lambda) := \int_{X} \e^{\lambda g} \;\d \mu$ is the \emph{partition function}, and the quantity $\Psi(\lambda) := \log Z(\lambda)$ is the \emph{generalized free energy}.  Note that, if $G$ is the macroscopic average energy, then $-k_{\mathrm{B}} T \: \Psi(-1/(k_{\mathrm{B}} T))$ is the familiar Helmholtz free energy at temperature $T$.

Denote by $\theta$ the map $\theta: [-\infty, +\infty] \rightarrow Y$ which associates a given $\lambda$ with a certain value of $y$ and is defined by $\theta(\lambda) := \int_{X} g \;\d P_{\lambda}$ for $\lambda \in \mathbb{R}$, $\theta(-\infty) := \ymin$, and $\theta(+\infty) := \ymax$.  The following lemmas will be needed.  The proofs are deferred to \App \ref{sec:GCL}.

\begin{lem}
\label{lem:FEL} % Free Energy Lemma
If\/ $\Psi''(\lambda)$ is nonzero for all\/ $\lambda \in (-\infty, +\infty)$, then the map\/ $\theta$ is invertible. Furthermore,\/ $y < y_*$ iff\/ $\lambda < 0$,\/ $y = y_*$ iff\/ $\lambda = 0$,\/ and $y > y_*$ iff\/ $\lambda > 0$.
\end{lem}

\begin{lem}
\label{lem:GIL} % Gibbs Infimum Lemma
Given\/ $y \in \interior{Y}$, let\/ $A_{\delta} = E_{g}^{-1}(B_{\delta})$ and\/ $\lambda = \theta^{-1}(y)$.  Then\/ $I_{\mu}(P_{\lambda}) = \inf I_{\mu}(\closure{A_{\delta}}) < \infty$ and\/ $I_{\mu}(P_{\lambda}) < I_{\mu}(P)$ for all\/ $P \in \closure{A_{\delta}} \setminus \set{P_{\lambda}}$,\/ where $\lambda = \theta^{-1}(y)$.
\end{lem}

\begin{lem}
\label{lem:CSL} % Continuity Set Lemma
Given\/ $y \in \interior{Y}$, the set\/ $A_{\delta} = E_{g}^{-1}(B_{\delta})$ is an\/ $I_{\mu}$-continuity set.
\end{lem}

Using the above lemmas we may deduce that, given $y \in \interior{Y}$, $A_{\delta} = E_{g}^{-1}(B_{\delta})$ is such that $\mu \not\in \boundary{A_{\delta}}$ and, for $\lambda = \theta^{-1}(y)$, $P_{\lambda}$ is the unique measure in $\closure{A_{\delta}}$ such that $I_{\mu}(P_{\lambda}) = \inf I(\closure{A_{\delta}}) = \inf I(\interior{A_{\delta}}) < \infty$. By \Thm \ref{thm:LDCC} we conclude that for any Borel set $A \subseteq \mathcal{P}(X)$,
\begin{equation}
\label{eqn:EMCPCN} % Empirical Measure Convergence in Probability, Conditioned on Nonequilibrium
  \lim_{n \rightarrow \infty} \mu^n[\set{L_n \in A} \:| \set{L_n \in A_{\delta}}] =
  \left\{
  \begin{array}{ll}
    0, & \mbox{if $P_{\lambda} \not\in \closure{A}$,} \\
    1, & \mbox{if $P_{\lambda} \in \interior{A}$.} \\
  \end{array}
  \right.
\end{equation}
In particular, for any Borel set $B \subset Y$,
\begin{equation}
\label{eqn:MACPCN} % Macroscopic Average Convergence in Probability, Conditioned on Nonequilibrium
  \lim_{n \rightarrow \infty} \mu^n[\set{G_{t} \in B} \:| \set{G \in B_{\delta}}] =
  \left\{
  \begin{array}{ll}
    0, & \mbox{if $\psi_{t}(y) \not\in \closure{B}$,} \\
    1, & \mbox{if $\psi_{t}(y) \in \interior{B}$,} \\
  \end{array}
  \right.
\end{equation}
where
\begin{equation}
\label{eqn:DC} % Deterministic Curve
\psi_{t}(y) := \int_{X} g \circ \varphi_{t} \;\d P_{\theta^{-1}(y)}.
\end{equation}
Note that $\psi_{0}(y) = y$ since $\varphi_{0}$ is the identity on $X$.  Conditioning on $\set{G_{t_0} \in B_{\delta}}$ merely shifts the time axis, in which case $G_{t}$ converges to $\psi_{t-t_0}(y)$ in probability.

%%%%%%%%%%%%%%%%%%%%%%%%%%%%%%%%%%%%%%%%%%%%%%%%%%%%%%%%%%%%%%%%%%%%%%%%%%%%

\section{The Deterministic Curve}
\label{sec:DC}

We have shown that, conditioned on $G \in B_{\delta}$, the macrostate $G_{t}$ at a given time, $t$, converges in probability to the expectation value $\psi_t(y)$, where $y$ is contained in $B_{\delta}$. Hence, of the initial microstates consistent with the initial macrostate, $y$, ``most'' will be such that the actual macrostate realized will be near this value. Now, each microstate, $(x_1, \ldots, x_n)$, gives rise to a collection, $\set{G_{t}(x_1, \ldots, x_n) : t \in T}$, of macrostates constituting a single trajectory. Likewise, each macrostate, $y$, gives rise to a collection, $\set{\psi_t(y) : t \in T}$, of expected macrostates, which we shall call the \emph{deterministic curve}. This graph represents the asymptotically deterministic behavior of the macrostate. In this section we shall investigate in what sense the deterministic curve is representative of a typical trajectory and consider several properties of the deterministic curve itself, considered as a function of time.

In general, there may be striking qualitative differences between the deterministic curve and a particular macrostate trajectory.  Suppose $\mu$ is invariant under $\varphi_t$ and $\mu^n[G^{-1}(B_{\delta})] > 0$.  The Poincar\'{e} recurrence theorem tells us that for some unbounded sequence $\tau_1, \tau_2, \ldots$ of times
\begin{equation}
\mu^n[\set{G_{\tau_1} \in B_{\delta},\: G_{\tau_2} \in B_{\delta},\: \ldots} \:| \set{G \in
B_{\delta}}] =  1,
\end{equation}
i.e.\/, the macrostate returns infinitely often to a neighborhood of its initial value, $y$, almost surely. Now suppose the map $t \mapsto \psi_t(y)$ has an attracting set $A$ with domain of attraction $D$ and take $B_{\delta} \subseteq D \setminus U$, where $U$ is a neighborhood of $A$.  For all $t$ sufficiently large, $\psi_t(y)$ will be in $U$, but $G_{t}$ will almost surely fall outside $U$ on the recurrence times $\tau_1, \tau_2, \ldots$. These recurrence times will depend upon $B_{\delta}$ and $\varphi_t$, of course, but typically increase rapidly with $n$. On a time scale small compared to $\tau_1$, one then expects the deterministic curve to be quite representative of a typical trajectory.  On time scales larger than $\tau_1$, however, the deterministic curve will be qualitatively quite different from a typical trajectory; the former converges to an attracting set while the latter exhibits quasi-periodic behavior.  This highlights the importance of a clearer understanding of the correspondence between the very distinct $n < \infty$ and $n = \infty$ cases.

%---------------------------------------------------------------------------

\subsection{Convergence to the Deterministic Curve}
\label{ssec:CDC}

Consider a finite set, $\set{t_1, \ldots, t_m}$, of times and for each time $t_i$ let $B_i \subseteq Y$ be a Borel set.  The set of microstates in which $G_{t_i} \in B_i$ for all $i$ is given by
\[
\bigcap_{i=1}^{m} \set{G_{t_i} \in B_i} = \set{ L_n \in \bigcap_{i=1}^{m} A_i},
\]
where $A_i = E_{g \circ \varphi_{t_i}}^{-1}(B_i)$.  According to \Eqn(\ref{eqn:EMCPCN}), with $A = \bigcap_{i=1}^{m} A_i$, the limiting conditional probability will be zero if $P_{\lambda} \not\in \closure{\bigcap_{i=1}^{m} A_i}$, where $\lambda = \theta^{-1}(y)$, and it will be one if $P_{\lambda} \in \interior{\left(\bigcap_{i=1}^{m} A_i\right)}$. Since the intersection is over a finite number of sets, we may use the fact that $\interior{\left(\bigcap_{i=1}^{m} A_i\right)} = \bigcap_{i=1}^{m} \interior{A_i}$ and $\closure{\bigcap_{i=1}^{m} A_i} \subseteq \bigcap_{i=1}^{m} \closure{A_i}$ to obtain the following:
\begin{equation}
\lim_{n \rightarrow \infty} \mu^{n}\left[\left. \bigcap_{i=1}^{m} \set{G_{t_i} \in B_i} \;\right| \set{G \in B_{\delta}} \right]
=
  \left\{
  \begin{array}{ll}
    0, & \mbox{if $\psi_{t_j}(y) \not\in \closure{B}_j$ for some $j$,} \\
    1, & \mbox{if $\psi_{t_i}(y) \in \interior{B}_i$ for all $i$.} \\
  \end{array}
  \right.
\end{equation}

For a countably infinite set of times, we may again conclude that the conditional probability goes to zero if $\psi_{t_j}(y) \not\in \closure{B_j}$ for some value of $j$, since $\bigcap_{i=1}^{\infty} \set{G_{t_i} \in B_i} \subseteq \set{G_{t_j} \in B_j}$. However, even if $\psi_{t_i}(y) \in \interior{B_i}$ for all $i$ and hence $P_{\lambda} \in \bigcap_{i=1}^{\infty} \interior{A_i}$, this does not assure us that $P_{\lambda} \in \interior{\left(\bigcap_{i=1}^{\infty} A_i\right)}$.  For the latter to be true, there must be a neighborhood of $P_{\lambda}$ which is sufficiently small so that it is contained in every $\interior{A_i}$.  Let us consider, then, conditions for which this is true.

Suppose each $B_i$ is an open interval of radius $\delta'$ centered at $\psi_{t_i}(y)$.  Now, each corresponding $A_i$ is an open set of probability measures which give an expectation of $g \circ \varphi_{t_i}$ in the open ball $B_i$.  Loosely speaking, if $g \circ \varphi_{t_i}$, and hence $E_{g \circ \varphi_{t_i}}$, varies too rapidly, then $A_i$ will be small as a result.  If, therefore, the functions $\set{g \circ \varphi_{t_i}}_{i \in \mathbb{N}}$ are somehow limited in how rapidly they may vary, then one might expect the sizes of $A_1, A_2, \ldots$ to be bounded from below, thus giving a nonempty interior for their intersection.  For a general metric space, $(X,d)$, one way to characterize how rapidly a function, $f$, may vary by its Lipschitz norm, $\|f\|_{\mathrm{L}}$, define as follows:
\begin{equation}
\|f\|_{\mathrm{L}} := \sup_{x \neq x'} \frac{|f(x)-f(x')|}{d(x, x')}.
\end{equation}
Differentiable functions will be Lipschitz if they have bounded derivatives; discontinuous functions, such as indicator functions, have infinite Lipschitz norm provided $d(x,x')$ may be made arbitrarily small.

To ensure well-defined expectations we require the functions to be bounded as well, so consider instead the bounded Lipschitz norm, $\|\cdot\|_{\mathrm{BL}}$, defined simply by $\|f\|_{\mathrm{BL}} := \|f\|_{\mathrm{L}} + \|f\|_{\infty}$.  (For a discussion of this norm, see Dudley \cite{Dudley}).  A set, $\set{f_i}_{i \in \mathbb{N}}$, of functions will be called \emph{uniformly bounded Lipschitz} if there exists a number, $K$, such that $\|f_i\|_{\mathrm{BL}} \le K$ for all $i \in \mathbb{N}$.  The following theorem states that for such observables the macroscopic trajectories converge in conditional probability to the deterministic curve on any countable set of times.

\begin{thm}
\label{thm:CIC} % Countably Infinite Convergence
If $\set{g \circ \varphi_{t_i}}_{i \in \mathbb{N}}$ are uniformly Lipschitz bounded functions, then for any $t_1, t_2, \ldots$,
\begin{equation}
\lim_{n \rightarrow \infty} \mu^{n}[ \set{\,|G_{t_i} - \psi_{t_i}(y)| < \delta',\, \forall i \in \mathbb{N}} \:| \set{G \in B_{\delta}}] = 1.
\end{equation}
\end{thm}

As noted above, this result may be proven by finding an open ball, $B_{\rho}(P_{\lambda}, \varepsilon)$, about $P_{\lambda}$ which is contained in every $A_i$.  Consider an arbitrary probability measure, $P$, in  $B_{\rho}(P_{\lambda}, \varepsilon)$ and observe that, since we have assumed $\|g \circ \varphi_{t_i}\|_{\mathrm{BL}} \le K$,
\begin{eqnarray*}%
|E_{g \circ \varphi_{t_i}}(P) - \psi_{t_i}(y)| &\le& \sup \set{|E_{f}(P) - E_{f}(P_{\lambda})| : \|f\|_{\mathrm{BL}} \le K} \\
&=& K \sup \set{|E_{f}(P) - E_{f}(P_{\lambda})| : \|f\|_{\mathrm{BL}} \le 1} \\
&\le& 2 K \rho(P, P_{\lambda}) < 2 K \varepsilon,
\end{eqnarray*}%
where the last inequality follows from \cite[pp.\ 310, 322]{Dudley}, since $(X,d)$ is a separable metric space.  Taking $\varepsilon \le \delta'/(2 K)$ shows that $E_{g \circ \varphi_{t_i}}(P) \in B_i$ for all $P \in B_{\rho}(P_{\lambda}, \varepsilon)$ and hence that $B_{\rho}(P_{\lambda}, \varepsilon) \subseteq A_i$.  Since $\varepsilon$ is independent of $i$, we conclude that $P_{\lambda} \in B_{\rho}(P_{\lambda}, \varepsilon) \subseteq \interior{\left(\bigcap_{i=1}^{\infty} A_i\right)}$.

For discrete time maps with suitable observables, \Thm \ref{thm:CIC} shows that the macrostate trajectories converge in conditional probability \emph{everywhere} to the deterministic curve.  This result may seem surprising, since we have seen that the long-time behavior of a typical trajectory may differ radically from that of the deterministic curve.  However, this simply means that the time scale on which the macroscopic behavior appears deterministic grows rapidly with the number of particles.

The case in which the set of times, $T$, takes on a continuum of values is somewhat more problematic owing to the fact that an arbitrary intersection of measurable sets need not be measurable.  If, however, $\set{G_{t} : t \in T}$ is sample continuous on an interval, $T$, i.e.\ if $t \mapsto G_{t}(x_1, \ldots, x_n)$ is continuous for every $(x_1, \ldots, x_n)$, then we may consider $\set{G_{t} : t \in T}$ to be a random process on the metric space of bounded continuous functions with the supremum norm.  For physical observables this is quite reasonable to suppose.  Much as in the theory of Brownian motion, we may then consider events of the form $\set{\sup_{t \in T} |G_{t} - \psi_t(y)| \ge \delta'}$ and their corresponding conditional probabilities as $n \rightarrow \infty$.  It may then be possible to derive a large deviation principle on the set of bounded continuous functions on $T$, much as is done for Brownian motion in Schilder's theorem \cite{Dembo_and_Zeitouni}, but here we do not pursue this matter further.

%---------------------------------------------------------------------------

\subsection{Properties of the Deterministic Curve}
\label{ssec:PDC}

In the previous section we considered the probabilistic convergence of macrostate trajectories to the asymptotic deterministic curve.  In this section we consider properties of the deterministic curve, $\set{\psi_t(y) : t \in T}$ as a function of $t$ in its own right.  Of course, we do not expect these properties to necessarily carry over to those of typical trajectories.  Nevertheless, they do give a clue to the behavior of these trajectories for large $n$ and relatively small $t$.

% Continuity

If $g$ is a discontinuous function, then specific realizations of $\set{G_{t} : t \in T}$ will also be discontinuous.  Since $G$ is the average of a bounded function, however, the size of these discontinuities will vanish as $n \rightarrow \infty$.  It is then reasonable to suppose that the deterministic curve, $\set{\psi_t(y) : t \in T}$, will be continuous in $t$.

Recall that $g$ and $\varphi_t$ are continuous $\mu$-almost everywhere.  Since $g$ is bounded $\mu$-a.e.\/, clearly $g \circ \varphi_t$ is so as well.  If we suppose $t \mapsto g(\varphi_t(x))$ is continuous for $\mu$-a.e.\ $x \in X$, then of course $g(\varphi_{t'}(x)) \rightarrow g(\varphi_t(x))$ as $t' \rightarrow t$ for $\mu$-a.e.\ $x \in X$.  By \Eqn~(\ref{eqn:DC}) and Lebesgue dominated convergence, this implies
\begin{equation}
\lim_{t' \rightarrow t} \psi_{t'}(y) = \lim_{t' \rightarrow t} \int_X g \circ \varphi_{t'} \;\d P_{\lambda} = \int_X g \circ \varphi_t \;\d P_{\lambda} = \psi_t(y),
\end{equation}
where $\lambda = \theta^{-1}(y)$.  Thus, $t \mapsto \psi_t(y)$ will be continuous provided $y \in \interior{Y}$ and $t \mapsto g(\varphi_t(x))$ is continuous for $\mu$-a.e.\ $x \in X$.

Now suppose only that $t \mapsto \varphi_t(x)$ is continuous for $\mu$-a.e.\ $x \in X$ and that $\varphi_t$ is $\mu$-nonsingular, i.e.\ $\mu \circ \varphi_t^{-1} \prec \mu$, for all $t \in T$.  Since $g$ is continuous $\mu$-a.e.\/, there exists an open set, $A$, with null complement such that $g$, restricted to $A$, is continuous.  Furthermore, since $\varphi_t$ is continuous $\mu$-a.e.\/, there is a set, $B$, with null complement such that $t \mapsto \varphi_t(x)$ is continuous for all $x \in B$. Thus, if $x \in \varphi_t^{-1}(A) \cap B$, then the composite map $t \mapsto \varphi_t(x) \mapsto g(\varphi_t(x))$ is continuous.  Since $\varphi_t$ is $\mu$-nonsingular,
\[
\mu[X \setminus (\varphi_t^{-1}(A) \cap B)]
     \le \mu[X \setminus \varphi_t^{-1}(A)] + \mu[X \setminus B]
     = \mu[\varphi_t^{-1}(X \setminus A)] = 0.
\]
Thus, $t \mapsto g(\varphi_t(x))$ is continuous for $\mu$-a.e.\ $x \in X$, and we conclude the following:

\begin{thm}
If $t \mapsto \varphi_t(x)$ is continuous for $\mu$-a.e.\ $x \in X$ and $\varphi_t$ is $\mu$-nonsingular for all $t \in T$, then $t \mapsto \psi_t(y)$ is continuous for all $y \in \interior{Y}$.
\end{thm}

% Symmetry

For many physical systems, the dynamical law $\varphi_t$ is not only invertible but also time reversible in the sense that $\varphi_t^{-1} = \varphi_{-t}$.  In such cases, we shall say that $\varphi_t$ is \emph{time reversible\/}. Time reversibility often appears in physical systems which have the added property that $\varphi_t \circ R = R \circ \varphi_{-t}$ for some involution $R = R^{-1}$, a property referred to as \emph{time reversal invariance\/}.  When such an $R$ exists, every particle microstate $x$ has a mirror point $R(x)$ such that $\varphi_t(x) = R(\varphi_{-t}(R(x)))$; hence, the trajectory of $x$ is mirrored by the trajectory of $R(x)$ with the direction of time reversed. We may then partition $X$ into disjoint sets $X_0 = R(X_0)$, $X_1$ and $R(X_1)$. If the observable and \textit{a priori\/} measure are invariant under $R$, i.e.\ $g = g \circ R$ and $\mu \circ R^{-1} = \mu$, then a typical initial microstate $(x_1, \ldots, x_n) \in \set{G \in B_{\delta}}$ will include roughly equal numbers of points from $X_1$ and $R(X_1)$. On this basis, one expects the trajectories $\set{G_{t}(x_1, \ldots, x_n) : t \ge 0}$ and $\set{G_{t}(x_1, \ldots, x_n) : t \le 0}$ to be similar, though not identical, when $n$ is large.  This suggests that the deterministic curve should be perfectly symmtric in time.  Indeed, it is easy to see that, if $\varphi_t$ is time reversal invariant under $R$ and both $g$ and $\mu$ are invariant under $R$, then
\begin{eqnarray*}%
\psi_t(y)
&=& \int_{X} (g \circ R \circ \varphi_t) \frac{\e^{\lambda g \circ R}}{Z(\lambda)} \;\d \mu \\
&=& \int_{X} (g \circ \varphi_{-t} \circ R) \frac{\e^{\lambda g \circ R}}{Z(\lambda)} \;\d (\mu \circ R^{-1}) \\
&=& \psi_{-t}(y).
\end{eqnarray*}%
Thus, time reversal invariance is sufficient for time symmetry of the deterministic curve, provided both the observable and the \textit{a priori} measure are invariant under $R$.

Suppose $g$ is a simple function of the form $g = \sum_{i=1}^{m} a_{i} 1_{C_i}$, where each $a_i$ is distinct and $C_1, \ldots, C_m$ form a partition of $X$.  Now, the deterministic curve for this $g$ is given by
\begin{equation}
\psi_t(y) = \sum_{i=1}^{m} \sum_{j=1}^{m} a_i \e^{\lambda a_j} \mu[\varphi_t^{-1}(C_i) \cap C_j] \;/\; \sum_{k=1}^{m} \e^{\lambda a_k} \mu[C_k],
\end{equation}
using \Eqn~(\ref{eqn:DC}).  Since we have supposed $g \circ R = g$, notice that $R^{-1}(C_i) = C_i$ for each $i$.  Furthermore, since $\mu$ is invariant under $R$,
\begin{eqnarray}%
\mu[\varphi_{t}^{-1}(C_i) \cap C_j]
&=& \mu[R^{-1}(\varphi_{t}^{-1}(C_i)) \cap R^{-1}(C_j)] \nonumber \\
&=& \mu[\varphi_{t}(R^{-1}(C_i)) \cap C_j] \nonumber \\
&=& \mu[\varphi_{t}(C_i) \cap C_j].
\end{eqnarray}%
Furthermore, if $\mu$ is invariant under $\varphi_t$, then
\begin{equation}
\mu[\varphi_{t}^{-1}(C_i) \cap C_j] = \mu[C_i \cap \varphi_{t}(C_j)].
\end{equation}
The system therefore exhibits \emph{strong detailed balance} in the sense that
\begin{equation}
\label{eqn:SDB} % Strong Detailed Balance
\mu[\varphi_t^{-1}(C_i) \cap C_j] = \mu[C_i \cap \varphi_t^{-1}(C_j)] \;\;\;\mbox{for all $i$, $j$, and $t$}.
\end{equation}

Conversely, if we suppose only that $\varphi_t^{-1} = \varphi_{-t}$ exists and preserves $\mu$, then \Eqn~(\ref{eqn:SDB}) implies time symmetry of the deterministic curve when $g$ is simple.  For example, suppose $g = a_1 1_{C_1} + a_2 1_{C_2}$ has only two possible states, with $C_1 = C$ and $C_2 = X \setminus C$.  For any $\varphi_t$ which preserves $\mu$,
\begin{eqnarray*}%
\mu[\varphi_t^{-1}(X \setminus C) \cap C]
&=& \mu[C] - \mu[\varphi_t^{-1}(C) \cap C] \\
&=& \mu[\varphi_t^{-1}(C)] - \mu[C \cap \varphi_t^{-1}(C)] \\
&=& \mu[(X \setminus C) \cap \varphi_t^{-1}(C)].
\end{eqnarray*}%
Thus, $g$ exhibits strong detailed balance. The corresponding deterministic curve will be time symmetric provided $\varphi_{t}^{-1} = \varphi_{-t}$ exists.  In general, macroscopic averages of two-state single-particle observables are always time symmetric, provided the dynamics are time reversible and preserve the \textit{a priori} measure.  A system which exhibits strong detailed balance need not, however, be time reversal invariant.  (Consider $C = [0,1]$ and $\varphi_t(x) = x+t$ on $\mathbb{R}$.  The only possible $R$ is $R(x) = -x$, yet clearly $R(C) \neq C$.)  Thus, time reversal invariant systems form a proper subset of all systems exhibiting time symmetry.

% Mixing and Convergence

Finally, let us consider the asymptotic behavior of the deterministic curve for large $t$.  Suppose once more that $\mu$ is invariant under $\varphi_{t}$. The collection, $\set{\varphi_t : t \in T}$,  of maps is said to be \emph{mixing} with respect to $\mu$ if, for any measurable subsets $A$ and $B$ of $X$,
\begin{equation}
\lim_{t \rightarrow \infty} \mu[\varphi_t^{-1}(A) \cap B] = \mu[A] \; \mu[B].
\end{equation}
From this definition and the fact that $P_{\lambda} \prec \mu$, it follows that \cite[p.\ 72]{Lasota_and_Mackey}
\begin{equation}
\lim_{t \rightarrow \infty} \int_{X} g \circ \varphi_t \;\d P_{\lambda} = \int_{X} g
\;\d \mu = y_*.
\end{equation}
If $\varphi_t$ is time reversible, then this result clearly holds in the limit $t \rightarrow -\infty$ as well. By its definition, mixing is both a necessary and sufficient condition for $\psi_t(y)$ to converge to $E_{g}(\mu) = y_*$ for \emph{any\/} $g$. In cases where the dynamics are not mixing, however, one may still have convergence in time for a restricted set of macroscopic functions.

As we have seen in our discussion of Poincar\'{e} recurrence, convergence in time for $\psi_t(y)$ need not imply convergence in time for $G_{t}$. In fact, such behavior is often quite unlikely. Nevertheless, on a short enough time scale, a scale which increases with $n$, and for large enough $n$, both $\psi_t(y)$ and $G_{t}$ will appear to converge to the same limit along the same trajectory.

%%%%%%%%%%%%%%%%%%%%%%%%%%%%%%%%%%%%%%%%%%%%%%%%%%%%%%%%%%%%%%%%%%%%%%%%%%%%

\section{Fractional Occupations}
\label{sec:FO}

Consider the case $g = 1_{C}$, for which $G_{t}$ is the fraction of points in $C$ at time $t$.  Since $\e^{\lambda g} = 1_{X \setminus C} + \e^{\lambda}1_{C}$, the corresponding partition function is
\begin{equation}
\label{eqn:FOPF} % Fractional Occupation Partition Function
Z(\lambda) = \mu[X \setminus C] + \e^{\lambda}\mu[C].
\end{equation}
Recall that $\theta(\lambda) = \Psi'(\lambda) = \e^{\lambda} \mu[C] / Z(\lambda)$ for $\lambda \in (-\infty, \infty)$; thus, for $y \in (0,1)$,
\begin{equation}
\label{eqn:FOTI} % Fractional Occupation Theta Inverse
\theta^{-1}(y) = \log \left[ \frac{y}{\mu[C]} \frac{\mu[X \setminus C]}{1-y}
\right].
\end{equation}
Using \Eqns (\ref{eqn:DC}), (\ref{eqn:FOPF}), and (\ref{eqn:FOTI}), we find
\begin{equation}
\label{eqn:FODC} % Fractional Occupation Deterministic Curve
\psi_{t}(y) = y \: \mu[\varphi_{t}^{-1}(C) \:| C] + (1-y) \:
\mu[\varphi_{t}^{-1}(C) \:| X \setminus C].
\end{equation}
This result is easily understood as follows:  The expected number of points in $C$ at time $t$ will be the number of points starting in $C$ times the fraction of those points expected to be in $C$ at time $t$ plus the number initially outside $C$ times the fraction of those points expected to be in $C$ at time $t$.

The derivation of \Eqn (\ref{eqn:FODC}) was valid for $y \in (0,1)$; using \Eqn (\ref{eqn:ECM}) we can see that it is valid for $y \in \set{0, 1}$ as well. Consider conditioning on $\set{G=1}$ and note that this is equivalent to conditioning on $C^{n}$, since $G(x_1, \ldots, x_n)=1$ iff $x_i \in C$ for all $i$. The conditional distribution of $G_{t}$ is therefore binomial with parameters $n$ and $\mu[\varphi_{t}^{-1}(C) \:| C]$. By the strong law of large numbers, $G_{t}$, conditioned on $\set{G=1}$, converges almost surely, hence in probability, to $\mu[\varphi_{t}^{-1}(C) \:| C]$.  A similar argument shows that $G_{t}$, conditioned on $\set{G=0}$, converges to $\mu[\varphi_{t}^{-1}(C) \:| X \setminus C]$.  We note in passing that the distribution of $\sqrt{n} G_{t}$ is asymptotically Gaussian by the central limit theorem, provided these transition probabilities are neither 0 nor 1.  Thus, deviations from the deterministic curve go to zero as $1/\sqrt{n}$ when $y \in \set{0,1}$.

\begin{figure}[hp]
\begin{center}
\scalebox{0.5}{\includegraphics{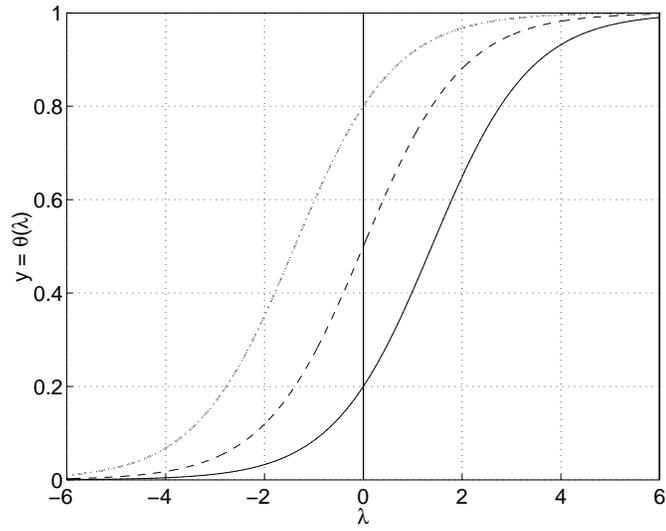}}
\end{center}
\caption{Three plots of $y = \theta(\lambda)$, where $g = 1_{C}$.  The value at $\lambda = 0$ is given by $\mu[C]$.}
\label{fig:theta}
\end{figure}

In \Fig \ref{fig:theta} we have plotted $\theta$ versus $\lambda$ for three different values of $\mu[C] = \theta(0)$. The strict monotonicity of the curve implies $\theta$ is invertible, in accordance with \Lem \ref{lem:FEL}. If, for example, $g = 1_{C}$ is the energy of a two state particle, with energies 0 and 1, then $\lambda = -(k_{\mathrm{B}} T)^{-1}$, where $T$ is the absolute temperature and $k_{\mathrm{B}}$ is Boltzmann's constant. An initial macrostate with energy density, $y$, near zero corresponds to a heavily populated low energy state and hence a low, positive temperature ($\lambda \ll 0$). If $y = \mu[C]$, then the initial macrostate is at its \textit{a priori\/} most likely state; if $\mu$ is invariant under $\varphi_t$, this corresponds to macroscopic equilibrium. Note that $\lambda \rightarrow 0^-$ corresponds to $T \rightarrow +\infty$; at high temperatures the particles are uniformly distributed, with respect to $\mu$, between the two energy states.  For macrostates beginning above equilibrium, i.e.\ $y > \mu[C]$, there is an effective population inversion similar to that found in systems of weakly coupled magnetic dipoles. Systems with small negative temperatures ($\lambda \gg 0$) tend to have more densely populated high energy states, while large negative temperatures ($\lambda \to 0^{+}$) again correspond to near equilibrium conditions. The single parameter $\mu[C]$ determines the asymmetry between states below equilibrium and those above. Thus, we see that $\lambda$ plays a role in defining initial macrostates which is analogous to that of temperature in defining equilibrium states.

Notice that the function $\theta$ is associated only with the initial macrostate.  The time-evolved behavior of fractional occupations is contained in the two transition probabilities in \Eqn (\ref{eqn:FODC}).  In general, these may be difficult to determine.  For the baker map, $\phi$, a discrete time map on the unit square \cite{Arnold_and_Avez}, these may be computed for rectangular cells with Lebesgue measure \cite{LaCour1999}.  This allows one to calculate $\psi_t(y)$ for several time iterations and to compare this with Monte Carlo simulations.  Although an abstract map, the baker map shares many of the relevant features of more realistic Hamiltonian dynamical systems.  In particular, it is Lebesgue measure preserving, mixing, and time reversal invariant in the sense that $\phi \circ R = R \circ \phi^{-1}$ for $R(x,y)=(y,x)$.

\begin{figure}[hp]
\begin{center}
\scalebox{0.5}{\includegraphics{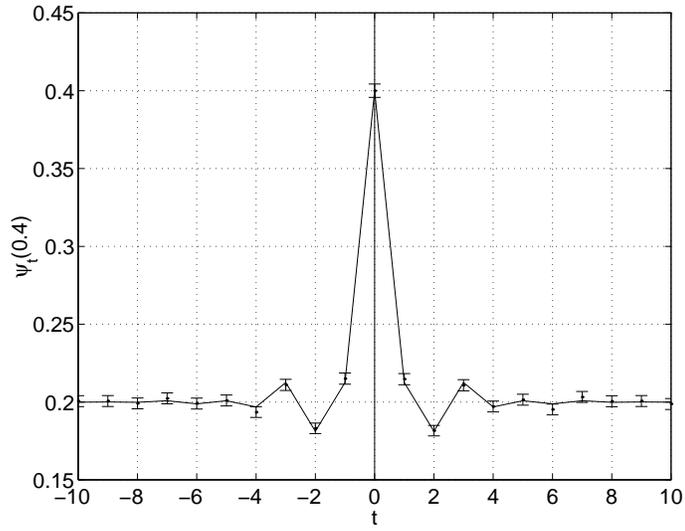}}
\end{center}
\caption{Plot of the fractional occupation of $C$ versus time for the baker map with $C = [0.2,\, 0.6) \times [0.0,\, 0.5)$ and $y=0.4$.  Straight lines are drawn between the values of $\psi_t(y)$ for each integer value of the iteration time $t$.  The solid dots are the values of $G_{t}$ for a single realization of an ensemble of $n = 50,000$ points.  The error bars are 95\% confidence intervals.}
\label{fig:baker}
\end{figure}

In \Fig \ref{fig:baker} we have plotted $\psi_t(y)$ for $t = -10, \ldots, 10$ and $y = 0.4$ using the baker map.  (Since the map is invertible, negative times refer to iterations of the inverse map.) The cell was chosen arbitrarily to be $C = [0.2,\, 0.6) \times [0.0,\, 0.5)$, for which $\mu[C] = 0.2$. The values of $\psi_t(y)$ are connected by straight solid lines in the figure. For comparison, a single realization of an ensemble of $n = 50,000$ points was generated which satiosfied the initial macrostate $y = 0.4$. This was done by drawing the first ${\lfloor n y \rfloor}$ points uniformly from $C$ and then drawing the rest from outside $C$.  (Here, ``uniformly'' means with respect to Lebesgue measure.) Once generated, the known form of the map $\varphi_t := \phi^t$ was used to time evolve the initial ensemble for each value of $t$. The fractional occupation, $G_{t}$, was then computed for each time-evolution of the initial ensemble and is indicated by a solid dot in the figure.

The qualitative behavior of $\psi_t(y)$ in \Fig \ref{fig:baker} is particularly notable in two regards.  First, it is readily observed that the plot is symmetric about $t = 0$; in particular, $\psi_{-t}(y) = \psi_{t}(y)$ exactly, while $G_{-t}$ and $G_{t}$ are only approximately equal.  This, as was shown in \Sec \ref{ssec:PDC}, is a general property of two-state systems for which $\varphi_t$ is $\mu$-measure preserving and time reversible.  Hence, there is no distinction between the forward and reverse time directions.  The second observation is that $\psi_t(y) \rightarrow \mu[C]$ as $t \rightarrow \pm \infty$, which is a direct consequence of the mixing property.  Thus, the baker map provides a simple model of an equilibrating macroscopic quantity.

A second comment is that, while at each \emph{given\/} time, $t$, the most probable macrostate is $\psi_t(y)$, for any finite $n$ the set $\set{\psi_t(y) : t \in \mathbb{Z}}$ is itself an improbable realization of $\set{G_{t} : t \in \mathbb{Z}}$. This may be understood by observing that, given $\varepsilon > 0$, we have $|\psi_t(y) - \mu[C]| < \varepsilon$ for all $|t|$ sufficiently large, yet, by Poincar\'{e} recurrence theorem, $|\psi_t(y) - \mu[C]| > \varepsilon$ for infinitely many values of $t$, almost surely.

The family of macroscopic maps, $\set{\psi_t: t \in \mathbb{Z}}$, does not form a group, or even a semigroup, in contrast to the family of microscopic maps, $\set{\varphi^t: t \in \mathbb{Z}}$.  Thus, while $x = \varphi_{-t}(\varphi_{t}(x))$, in general $y \neq \psi_{-t}(\psi_{t}(y)) =\psi_{t}(\psi_{t}(y))$, since $\psi_{t}$ is time symmetric.  Furthermore, $\set{\psi_{t}: t \ge 0}$ does not even form a semigroup, since this would imply $|\psi_{t}(y) - y_{*}| \ge |\psi_{t+s}(y) - y_{*}|$ for all $s, t \ge 0$, i.e.\ that all future macrostates are closer to equilibrium than their predecessors.  To understand this note that, while $\psi_{t+s}(y)$ describes the state of an observable at time $t+s$ whose value was $y$ at time zero, $\psi_{s}(\psi_{t}(y))$ describes the state of an observable at time $t+s$ whose value was $\psi_t(y)$ at time $t$.  The latter corresponds to a rerandomization of the original distribution, which removes correlations that would otherwise be preserved by the dynamics and causes disagreement with the actual time evolution of the observable.

\begin{figure}[hp]
\begin{center}
\scalebox{0.5}{\includegraphics{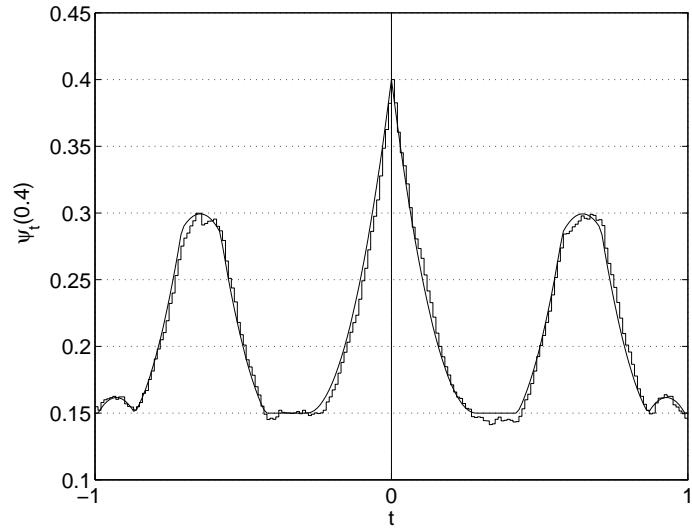}}
\end{center}
\caption{Plot of the expected fractional occupation of $C$ versus time for the ergodic rotation map with $C = [0.2,\, 0.6) \times [0.0,\, 0.5)$ and $y=0.4$.  A particular realization using $n = 5,000$ is plotted for comparison}
\label{fig:rotation}
\end{figure}

The baker map is a discrete time map, whereas the dynamics of physical systems are given by continuous time flows.  A simple example of this type of system is the rotation map on the unit square, which is given by
\begin{equation}
\varphi_t(x_1, x_2) = (x_1 + \omega_1 t,\: x_2 + \omega_2 t) \bmod 1.
\end{equation}
A plot of $\psi_t(y)$ for $\omega_1 = \sqrt{2}$, $\omega_2 = \sqrt{3}$, and $y = 0.4$ is given in \Fig \ref{fig:rotation}.  We again note the perfect symmetry about $t = 0$ found previously in the baker map. Unlike the baker map, however, the expectation value does not converge to an asymptotic value but varies quasi periodically in time. (The flat portions of the graph occur when $\varphi_t^{-1}(C) \cap C$ is empty and hence only $0.6 \times (0.2/0.8) = 0.15$ of the remaining points are expected to be in $C$.) A particular realization using $n = 5,000$ is plotted for comparison.

Near $t = 0$, the deterministic curve is linear with a slope pointing toward the equilibrium value, $y_* = 0.2$, as $|t|$ increases, a behavior which holds generally for any value of $y$.  Thus, the initial tendency of the system is to move monotonically toward the equilibrium value.  Such behavior has been ascribed to Boltzmann's H-function \cite{Ehrenfest}, though extrapolated to include times far from zero as well.  As we have seen from the maps considered here, this extrapolation need not be valid.  Since the H-function is computed from fractional occupations, though, it seems reasonable that monotonicity toward equilibrium should hold at least for $t$ near zero.

%%%%%%%%%%%%%%%%%%%%%%%%%%%%%%%%%%%%%%%%%%%%%%%%%%%%%%%%%%%%%%%%%%%%%%%%%%%%

\section*{Acknowledgements}

The authors would like to thank D. Driebe and H. Hasagawa for their helpful comments and constructive criticisms.

%%%%%%%%%%%%%%%%%%%%%%%%%%%%%%%%%%%%%%%%%%%%%%%%%%%%%%%%%%%%%%%%%%%%%%%%%%%%

\section{Discussion}

We have considered a general class of systems composed of identical constituents, here called ``particles,'' that are dynamically noninteracting.  For the microstates of the collective system, we supposed there is an \textit{a priori\/} measure, typically an invariant measure, that describes the distribution of these microstates in the absence of any restrictions based on the given macrostate.  We further supposed that the particles are statistically independent, that any correlations among them arise only by the need to satisfy the given macroscopic constraint.  No attempt was made to justify these assumptions at a more fundamental level, though we believe they are quite reasonable for many physical systems.

What we have shown is that the time evolution of a particular macroscopic variable, namely the average over certain real-valued single-particle functions, is such that it converges in a probabilistic sense to a well defined curve as the number of particles tends to infinity.  Specifically, we have derived a map $\psi_t$ such that, if the macrostate at time 0 is constrained to be near a value $y$, then the macrostate at time $t$ will be in a given neighborhood of $\psi_t(y)$ with a probability approaching one as the number of particles tends to infinity.  The map $\psi_t$ was defined in terms of an expectation with respect to a canonical distribution in which $y$ plays the role of an average energy in the familiar thermodynamic formalism.  The restrictions on the single-particle function were that it be bounded and continuous almost everywhere in a sense specified by the \textit{a priori\/} measure.  We found that the family of macroscopic maps, $\set{\psi_{t}: t \in T}$, in general forms neither a group nor a semigroup, even if the family of microscopic maps, $\set{\varphi_{t}: t \in T}$, has this property.

Having established this basic convergence result for a given time, we then considered how well the deterministic curve, the graph of $\psi_t(y)$ versus $t$, represented the behavior of a typical realization of the macrostates over all time.  We found that the two may differ qualitatively quite substantially; while there may be good agreement on a finite set of selected times, there will typically be times at which they differ substantially.  This was particularly true of mixing systems, for which $\psi_t(y)$ always converges in the long time limit, while a typical trajectory exhibits recurrences. Under some more restrictive conditions we proved convergence on any countably infinite set of times, but even then recurrences are possible when $n$ is finite.

With these caveats on the correspondence between the finite and infinite particles cases, we considered some general properties of the expectation curve as a function of time.  We found that, despite the fact that the macrostates may evolve discontinuously, the deterministic curve may be continuous in time.  We also found that, for systems which are time reversal invariant, the deterministic curve is symmetric in time about $t=0$, the point at which conditioning of initial macrostates takes place.  These properties were then related to familiar geometric properties attributed to Boltzmann's H-curve.

We have not considered extensions of these results to macroscopic variables in, say, $\mathbb{R}^d$, which would involve issues of convexity that make the extension nontrivial.  The general problem of interacting particles poses a  greater difficulty and requires a significant change of methodology, though we conjecture that similar results will hold if $\psi_t(y)$ is defined as a limit of $n$-particle expectations.

%%%%%%%%%%%%%%%%%%%%%%%%%%%%%%%%%%%%%%%%%%%%%%%%%%%%%%%%%%%%%%%%%%%%%%%%%%%%

\appendix

%%%%%%%%%%%%%%%%%%%%%%%%%%%%%%%%%%%%%%%%%%%%%%%%%%%%%%%%%%%%%%%%%%%%%%%%%%%%%%

\section{Proof of Theorem \ref{thm:LDCC}}
\label{sec:LDCC}

\begin{pf}
It suffices to consider $x_{B} \not\in \closure{A}$ since, if $x_{B} \in \interior{A}$, then $x_{B} \not\in X \setminus \interior{A} = \closure{X \setminus A}$.

Since $x_* \notin \boundary{B}$, either $x_* \in \interior{B}$ or $x_* \not\in \closure{B}$.  Suppose the former.  By \Eqn(\ref{eqn:LDZOC}), $P_{n}[B] \rightarrow 1$  as $n \rightarrow \infty$, which implies $P_{n}[A \:| B] \rightarrow \lim_{n \rightarrow \infty} P_ {n}[A]$. Now, $P_{n}[A] \rightarrow 0$ if $x_* \not\in \closure{A}$, while $P_{n}[A] \rightarrow 1$ if $x_* \in \interior{A}$. Since $x_{B} = x_*$ for this case, the result is proven.

Since $x_* \not\in \closure{B}$, $0 < \inf I(\closure{B}) \le \inf I(\interior{B})$.  By \Eqn(\ref{eqn:LDULB}) we have, for any $\varepsilon > 0$,
\[
P_{n}[B] > \exp[-a_{n}(1+\varepsilon) \inf I(\interior{B})] > 0
\]
for all $n$ sufficiently large; thus, $P_{n}[A \:| B]$ is well defined.
 Suppose further that $\inf I(\closure{A \cap B}) < \infty$.  From \Eqn(\ref{eqn:LDULB}) we may also deduce that for all $n$ sufficiently large,
\begin{eqnarray*}
P_{n}[A \:| B]
&\le& \frac{\exp[{-a_{n}(1-\varepsilon)
\inf I(\closure{A \cap B})}]}{\exp[-a_{n}(1+\varepsilon) \inf
I(\interior{B})]} \nonumber \\
&=& \exp[-a_{n}(1-\varepsilon) \inf I(\closure{A \cap B}) + a_{n}(1+\varepsilon)
\inf I(\interior{B})].
\end{eqnarray*}

Now suppose $x_{B} \not\in \closure{A}$.  To show that $P_{n}[A \:| B] \rightarrow 0$, it will suffice to show that we may choose $\varepsilon$ such that
\[
(1-\varepsilon) \inf I(\closure{A \cap B}) - (1+\varepsilon) \inf
I(\interior{B}) > 0.
\]
Since this means we must choose $\varepsilon$ small enough so that
\[
\varepsilon < \frac{\inf I(\closure{A \cap B}) - \inf I(\interior{B})}{\inf
I(\closure{A \cap B}) + \inf I(\interior{B})},
\]
we see that it will suffice to show that $\inf I(\closure{A \cap B}) > \inf I(\interior{B})$.  (Note that, since $\inf I(\interior{B}) > 0$, the denominator in the above inequality is indeed nonzero.)

We have assumed $B$ is an $I$-continuity set, so $\inf I(\interior{B}) = \inf I(\closure{B}) = I(x_{B}) < \infty$. Now, if $\closure{A \cap B} = \varnothing$ then $\inf I(\closure{A \cap B}) = \infty > \inf I(\closure{B}) = \inf I(\interior{B})$ and we are done.  Suppose that $\closure{A \cap B} \neq \varnothing$. Then there exists an $x \in \closure{A \cap B}$ such that $I(x) = \inf I(\closure{A \cap B})$, since $I$ is a good rate function. Notice that $x_{B} \not\in \closure{A \cap B}$, since $\closure{A \cap B} \subseteq \closure{A} \cap \closure{B} \subseteq \closure{A}$ and $x_{B} \not\in \closure{A}$.  Clearly, then, $x_{B} \neq x$. Since $\closure{A \cap B} \subseteq \closure{B}$ as well, $\inf I(\closure{A \cap B}) \ge \inf I(\closure{B})$, or, equivalently, $I(x) \ge I(x_{B})$. Equality cannot hold, however, since, if that were the case, then $I(x)$ would equal $\inf I(\closure{B})$, in violation of the assumed uniqueness of $x_{B}$. Therefore, $\inf I(\closure{A \cap B}) > \inf I(\closure{B}) = \inf I(\interior{B})$.

Now suppose $\inf I(\closure{A \cap B}) = \infty$ instead.  By \Eqn(\ref{eqn:LDZOC}), this implies $P_{n}[A \cap B] \rightarrow 0$. Since $P_{n}[B] > 0$ for all $n$ sufficiently large, this implies $P_{n}[A \:| B] \rightarrow 0$.
\end{pf}

%%%%%%%%%%%%%%%%%%%%%%%%%%%%%%%%%%%%%%%%%%%%%%%%%%%%%%%%%%%%%%%%%%%%%%%%%%%%%

\section{Proof of Gibbs Conditioning Lemmas}
\label{sec:GCL}

The proofs of lemmas \ref{lem:FEL} and \ref{lem:GIL} are adapted from those of Ellis \cite{Ellis1999}, who considers the case in which $g$ is a simple function.  The extension to a general bounded measurable function is similar but not trivial.  Dembo and Zeitouni \cite[p. 294-7]{Dembo_and_Zeitouni} prove lemma \ref{lem:GIL} for the $\tau$-topology, for which $E_{g}$ is continuous for any bounded $g$.  The proof given here applies for the weaker Prohorov metric topology.

\subsection{Proof of Lemma \ref{lem:FEL}}
\label{ssec:FEL}

\begin{pf}
We first note that, since $\lambda \mapsto e^{\lambda g(x)}$ is bounded and differentiable to all orders for $\mu$-a.e.\ $x \in X$, by Lebesgue dominated convergence $\Psi'$ and $\Psi''$ are well-defined and continuous. Specifically, $\Psi'(\lambda) = \int_{X} g \;\d P_{\lambda}$ and $\Psi''(\lambda) = \int_{X} g^2 \;\d P_{\lambda} - (\int_{X} g \;\d P_{\lambda})^2 \ge 0$ for $\lambda \in \mathbb{R}$. By assumption $\Psi''(\lambda)$ is nonzero, so in fact $\Psi''(\lambda) > 0$ and hence $\Psi'$ increases monotonically.

We now show that $\Psi'(\lambda) \rightarrow \ymax$ as $\lambda \rightarrow \infty$.  For $\lambda > 0$, note that
\[
|\Psi'(\lambda) - \ymax| = \left| \frac{\int_{X} (g-\ymax) \e^{\lambda g} \d \mu}{\int_{X} \e^{\lambda g} \d \mu} \right| \le \frac{\int_{Y} |y-\ymax| \e^{\lambda y} \d \nu(y)}{\int_{Y} \e^{\lambda y} \d \nu},
\]
where $\nu := \mu \circ g^{-1}$.  Now, for any $\delta > 0$,
\begin{eqnarray*}%
|\Psi'(\lambda) - \ymax|
&\le& \frac{|\ymin-\ymax| \int_{\ymin}^{\ymax-\delta}
\e^{-\lambda(\ymax - y)} \d \nu(y)}{\int_{Y} \e^{-\lambda(\ymax - y)} \d
\nu} + \frac{\int_{\ymax-\delta}^{\ymax} |y-\ymax| \e^{\lambda y} \d
\nu(y)}{\int_{Y} \e^{\lambda y} \d \nu} \\
&\le& |\ymin-\ymax| \, \e^{-\lambda \delta} \, \nu[(\ymin, \ymax -\delta)] +
\frac{\delta \, \e^{\lambda \ymax} \, \nu[(\ymax-\delta,
\ymax)]}{\e^{\lambda(\ymax-\delta)} \, \nu[(\ymax-\delta, \ymax)]} \\
&\le& |\ymin-\ymax| \, \nu[(\ymin, \ymax)] \, \e^{-\lambda \delta} +
\delta \, \e^{\lambda \delta}.
\end{eqnarray*}%
For $\lambda > 1$, take $\delta = \lambda^{-1} \log(\log(\lambda))$ and note that $\e^{-\lambda \delta} = 1/\log(\lambda)$ and $\delta \, \e^{\lambda \delta} = \lambda^{-1} \log(\lambda) \log(\log(\lambda))$.  The latter two terms vanish as $\lambda \rightarrow \infty$, as may be readily verified by L'Hospital's rule.

A similar argument shows that $\Psi'(\lambda) \rightarrow \ymin$ as $\lambda \rightarrow -\infty$.  Continuity then implies that $\Psi'$ is surjective onto $Y$. Thus, $\theta$ is invertible with $\theta^{-1}(y) := (\Psi')^{-1}(y)$ for $y \in \interior{Y}$, $\theta^{-1}(\ymin) := -\infty$, and $\theta^{-1}(\ymax) := +\infty$. Since $\theta$ increases monotonically and $\theta(0) = y_*$, $\lambda > 0$ implies $y = \theta(\lambda) > y_*$, while $\lambda < 0$ implies $y = \theta(\lambda) < y_*$.  Since $\theta$ is invertible, we have conversely that $y > y_*$ implies $\lambda > 0$, while $y < y_*$ implies $\lambda < 0$. Clearly, $\lambda = 0$ if and only if $y = y_*$.
\end{pf}

\subsection{Proof of Lemma \ref{lem:GIL}}
\label{ssec:GIL}

\begin{pf}
Since $P_{\lambda} \prec \mu$ we have
\begin{eqnarray*}
  I_{\mu}(P_{\lambda}) &=& \int_{X} \log \frac{\d P_{\lambda}}{\d \mu} \;\d P_{\lambda}
    = \int_{X} \log \frac{\e^{\lambda g}}{Z(\lambda)} \;\d P_{\lambda} \\
  &=& \lambda \int_{X} g \;\d P_{\lambda} - \Psi(\lambda) = \lambda y - \Psi(\lambda).
\end{eqnarray*}
Since $g$ is bounded, $\Psi(\lambda) = \log \int_{X} \e^{\lambda g} \;\d \mu > -\infty$ and hence $I_{\mu}(P_{\lambda}) < \infty$.

Now let $P \in \closure{A_{\delta}} \setminus \set{P_{\lambda}}$.  If $P \not\prec \mu$, then $I_{\mu}(P) = \infty$ and $I_{\mu}(P_{\lambda}) < I_{\mu}(P)$. Consider then $P \prec \mu$. Using the chain rule, ${\d P}/{\d \mu} = ({\d P}/{\d P_{\lambda}}) ({\d P_{\lambda}}/{\d \mu})$ \cite[pp.\ 265-6]{Rao}, we find
\begin{eqnarray*}
I_{\mu}(P)
&=& \int_{X} \log \frac{\d P}{\d \mu} \;\d P = \int_{X} \log \frac{\d P}{\d P_{\lambda}} \;\d P + \int_{X} \log \frac{\d P_{\lambda}}{\d \mu} \;\d P \\
&=& I_{P_{\lambda}}(P) + \int_{X} \log \frac{\e^{\lambda g}}{Z(\lambda)} \;\d P = I_{P_{\lambda}}(P) + \lambda \int_{X} g \;\d P - \Psi(\lambda) \\
&>& \lambda \int_{X} g \;\d P - \Psi(\lambda),
\end{eqnarray*}
where we have used the fact that $I_{P_{\lambda}}(P) > 0$ since $P \neq P_{\lambda}$.

Since $P \prec \mu$, $E_{g}$ is continuous at $P$; thus, $P \in \closure{A_{\delta}}$ implies $E_{g}(P) = \int_{X} g \;\d P \in \closure{B_{\delta}}$.  If $y < y_*$ then $\closure{B_{\delta}} = [y - \delta, y]$ and $\lambda < 0$ by \Lem \ref{lem:FEL}. Now, $\lambda < 0$ and $\int_{X} g \;\d P \le y$ imply $\lambda \int_{X} g \;\d P \ge \lambda y$. If, on the other hand, $y > y_*$, then $\closure{B_{\delta}} = [y, y + \delta]$ and $\lambda > 0$, while $\int_{X} g \;\d P \ge y$, so again $\lambda \int_{X} g \;\d P \ge \lambda y$. Finally, if $y = y_*$ then $\lambda = 0$ and $\lambda \int_{X} g \;\d P \ge \lambda y$ holds trivially. Thus, for all $P \in \closure{A_{\delta}}$,
\[
I_{\mu}(P_{\lambda}) = \lambda y - \Psi(\lambda) \le  \lambda \int_{X} g \;\d P - \Psi(\lambda) < I_{\mu}(P).
\]
Since $P_{\lambda} \in A_{\delta} \subseteq \closure{A_{\delta}}$, this completes the proof.
\end{pf}

\subsection{Proof of Lemma \ref{lem:CSL}}
\label{ssec:CSL}

\begin{pf}
Suppose $y \le y_*$ and let $\lambda_n = \lambda - 1/n$, where $\lambda = \theta^{-1}(y)$.  (A similar argument may be applied if $y \ge y_*$.)  Since $\theta$ is strictly monotonic, $E_{g}(P_{\lambda_n}) = \theta(\lambda_n) < \theta(\lambda) = y$.  We will first show that $E_{g}(P_{\lambda_n}) \in \interior{B_{\delta}}$ for all $n$ sufficiently large and that $E_{g}(P_{\lambda_n}) \rightarrow E_{g}(P_{\lambda})$.  Since $E_{g}$ is continuous at $P_{\lambda}$, it will suffice to show that $P_{\lambda_n} \rightarrow P_{\lambda}$ in $\rho$.

Now, convergence in $\rho$ is equivalent to the weak convergence of $P_{\lambda_n}$ to $P_{\lambda}$ \cite[p.\ 310]{Dudley}.  Thus, let $h$ be an arbitrary bounded continuous function on $X$ and define $F$ such that $F(\lambda') = \int_{X} h \; \e^{\lambda' g}/Z(\lambda') \;\d \mu$ for $\lambda' \in \mathbb{R}$.  Since the integrand is bounded and continuous for $\mu$-a.e.\ $x$, it follows that $F$ is continuous everywhere and $F(\lambda_n) \rightarrow F(\lambda)$.  This proves weak convergence and hence convergence in $\rho$.

We have shown that $E_{g}(P_{\lambda_n}) \in \interior{B_{\delta}}$ for all $n$ sufficiently large.  From this it follows that $P_{\lambda_n} \in \interior{A_{\delta}} \subseteq \closure{A_{\delta}}$ and hence \( I_{\mu}(P_{\lambda_n}) = \lambda_n E_{g}(P_{\lambda_n}) - \Psi(\lambda_n) > I_{\mu}(P_{\lambda}) \) for all $n$ sufficiently large.  Now, $\inf I_{\mu}(\interior{A_{\delta}}) \ge \inf I_{\mu}(\closure{A_{\delta}})$, so \(I_{\mu}(P_{\lambda_n}) \ge \inf I_{\mu}(\interior{A_{\delta}}) \ge \inf I_{\mu}(\closure{A_{\delta}}) = I_{\mu}(P_{\lambda})\).  Since $E_{g}(P_{\lambda_n}) \rightarrow E_{g}(P_{\lambda})$ and $\Psi(\lambda_n) \rightarrow \Psi(\lambda)$ as $\lambda_n \rightarrow \lambda$, it is clear that $I_{\mu}(P_{\lambda_n}) \rightarrow I_{\mu}(P_{\lambda})$.   Hence $\inf I_{\mu}(\interior{A_{\delta}}) = \inf I_{\mu}(\closure{A_{\delta}})$.
\end{pf}

%%%%%%%%%%%%%%%%%%%%%%%%%%%%%%%%%%%%%%%%%%%%%%%%%%%%%%%%%%%%%%%%%%%%%%%%%%%%

\bibliographystyle{hunsrt}
\bibliography{lacour}

\begin{thebibliography}{10}

\bibitem{vanKampen1961}
N.~G. van Kampen.
\newblock A power series expansion of the master equation.
\newblock {\em Canadian Journal of Physics}, 39:551, 1961.

\bibitem{Kurtz1972}
Thomas~G. Kurtz.
\newblock The relationship between stochastic and deterministic models for
  chemical reactions.
\newblock {\em Journal of Chemical Physics}, 57(7):2976--2978, 1972.

\bibitem{Kubo1973}
Ryogo Kubo, Kazuhiro Matsuo, and Kazuo Kitahara.
\newblock Fluctuations and relaxation of macrovariables.
\newblock {\em Journal of Statistical Physics}, 9(1):51--96, 1973.

\bibitem{Spohn}
Herbert Spohn.
\newblock {\em Large Scale Dynamics of Interacting Particles}.
\newblock Springer-Verlag, New York, 1991.

\bibitem{Davies}
Edward~B. Davies.
\newblock {\em The Theory of Open Systems}.
\newblock Academic Press, London, 1976.

\bibitem{Sklar}
Lawrence Sklar.
\newblock {\em Physics and Chance: Philosophical Issues in the Foundations of
  Statistical Mechanics}.
\newblock Cambridge University Press, Cambridge, 1993.

\bibitem{Mehra1972}
J.~Mehra and E.~C.~G. Sudarshan.
\newblock Some reflections on the nature of entropy, irreversibility and the
  second law of thermodynamics.
\newblock {\em Il Nuovo Cimento}, 11 B(2):215--256, 1972.

\bibitem{Ehrenfest}
Paul Ehrenfest and Tatiana Ehrenfest.
\newblock {\em The Conceptual Foundations of the Statistical Approach in
  Mechanics}.
\newblock Cornell University Press, Ithaca, NY, 1959.

\bibitem{Dudley}
Richard~M. Dudley.
\newblock {\em Real Analysis and Probability}.
\newblock Chapman \& Hall, New York, 1989.

\bibitem{Ash}
Robert~B. Ash.
\newblock {\em Real Analysis and Probability}.
\newblock Academic Press, New York, 1972.

\bibitem{Boltzmann1877}
Ludwig Boltzmann.
\newblock {\"{U}}ber die beziehung zwischen dem zweiten hauptsatze der
  mechanischen w{\"{a}}rmetheorie und der wahrscheinlichkeitsrechnung
  respecktive den s{\"{a}}tzen {\"{u}}ber das w{\"{a}}rmegleichgewicht ({O}n
  the relationship between the second law of the mechanical theory of heat and
  the probability calculus).
\newblock {\em Wiener Berichte}, 2(76):373--435, 1877.

\bibitem{Einstein1907}
Albert Einstein.
\newblock {\em Annalen der Physik}, 22:180, 1907.

\bibitem{Deuschel_and_Stroock}
Jean-Dominique Deuschel and Daniel~W. Stroock.
\newblock {\em Large Deviations}.
\newblock Academic Press, San Diego, 1989.

\bibitem{Ellis}
Richard~S. Ellis.
\newblock {\em Entropy, Large Deviations, and Statistical Mechanics}.
\newblock Springer-Verlag, New York, 1985.

\bibitem{Dembo_and_Zeitouni}
Amir Dembo and Ofer Zeitouni.
\newblock {\em Large Deviations Techniques and Applications}.
\newblock Jones and Bartlett, Boston, 1993.

\bibitem{Dupuis_and_Ellis}
Paul Dupuis and Richard~S. Ellis.
\newblock {\em A Weak Convergence Approach to the Theory of Large Deviations}.
\newblock John Wiley \& Sons, New York, 1997.

\bibitem{Ruelle}
David Ruelle.
\newblock {\em Statistical Mechanics: Rigorous Results}.
\newblock Benjamin, New York, 1969.

\bibitem{Lanford1973}
Oscar E.~Lanford III.
\newblock Entropy and equilibrium states in classical statistical mechanics.
\newblock In A.~Lenard, editor, {\em Statistical Mechanics and Mathematical
  Problems}, volume~20 of {\em Lecture Notes in Physics}, pages 1--111.
  Springer-Verlag, 1973.

\bibitem{Kifer1990}
Yuri Kifer.
\newblock Large deviations in dynamical systems and stochastic processes.
\newblock {\em Transactions of the American Mathematical Society},
  321(2):505--524, 1990.

\bibitem{Ellis1999}
Richard~S. Ellis.
\newblock The theory of large deviations: {F}rom boltzmann's 1877 calculation
  to equilibrium macrostates in 2d turbulence.
\newblock {\em Physica D}, 1999.

\bibitem{Lasota_and_Mackey}
Andrzej Lasota and Michael~C. Mackey.
\newblock {\em Chaos, Fractals, and Noise}.
\newblock Springer-Verlag, New York, 1994.

\bibitem{Arnold_and_Avez}
Vladimir~I. Arnol'd and Andre Avez.
\newblock {\em Ergodic Problems of Classical Mechanics}.
\newblock Benjamin, New York, 1968.

\bibitem{LaCour1999}
Brian~R. {La Cour} and William~C. Schieve.
\newblock Quasi {M}arkovian behavior in mixing maps.
\newblock {\em Physica D}, 133(1--4):309--320, 1999, cond-mat/9902105.

\bibitem{Rao}
Malempati~M. Rao.
\newblock {\em Measure Theory and Integration}.
\newblock John Wiley \& Sons, New York, 1987.

\end{thebibliography}

%%%%%%%%%%%%%%%%%%%%%%%%%%%%%%%%%%%%%%%%%%%%%%%%%%%%%%%%%%%%%%%%%%%%%%%%%%%%

\newpage

\newpage

\newpage

%%%%%%%%%%%%%%%%%%%%%%%%%%%%%%%%%%%%%%%%%%%%%%%%%%%%%%%%%%%%%%%%%%%%%%%%%%%%

\end{document}